\documentclass[aps,prl,twocolumn,showpacs,superscriptaddress,groupedaddress,nofootinbib]{revtex4}  
\usepackage{graphicx}  
\usepackage{dcolumn}   
\usepackage{bm}       
\usepackage{amssymb}   
\usepackage[section]{placeins}
\usepackage{capt-of}
\usepackage{color}
\usepackage[fleqn]{amsmath}

\usepackage{natbib}

\begin{document}

\widetext
\title{Highly spin-polarized deuterium atoms from the UV dissociation of Deuterium Iodide.}

\author{Dimitris Sofikitis}
\affiliation{Institute of Electronic Structure and Laser, Foundation for Research and Technology-Hellas, 71110 Heraklion-Crete, Greece.}
\affiliation{Department of Physics, University of Crete, 71003, Heraklion-Crete, Greece.}

\author{Pavle Glodic}
\affiliation{Institute of Electronic Structure and Laser, Foundation for Research and Technology-Hellas, 71110 Heraklion-Crete, Greece.}

\author{Greta Koumarianou}
\affiliation{Institute of Electronic Structure and Laser, Foundation for Research and Technology-Hellas, 71110 Heraklion-Crete, Greece.}
\affiliation{Department of Chemistry, University of Crete, 71110, Heraklion-Crete, Greece.}

\author{Hongyan Jiang}
\affiliation{Max Planck Institute for Biophysical Chemistry Fassberg 11, D-37077 G{\"o}ttingen, Germany.}

\author{Lykourgos Bougas}
\altaffiliation[Current address:]{ Johannes Gutenberg-Universitat Mainz, 55128 Mainz, Germany}
\affiliation{Institute of Electronic Structure and Laser, Foundation for Research and Technology-Hellas, 71110 Heraklion-Crete, Greece.}
\affiliation{Department of Physics, University of Crete, 71003, Heraklion-Crete, Greece.}

\author{Peter C. Samartzis}
\affiliation{Institute of Electronic Structure and Laser, Foundation for Research and Technology-Hellas, 71110 Heraklion-Crete, Greece.}
\affiliation{Department of Chemistry, University of Crete, 71110, Heraklion-Crete, Greece.}

\author{Alexander Andreev}
\affiliation{IELI-ALPS, 6720 Szeged, Hungary}
\affiliation{Max Born Institute, 12489 Berlin, Germany.}

\author{T. Peter Rakitzis$^\star$}
\affiliation{Institute of Electronic Structure and Laser, Foundation for Research and Technology-Hellas, 71110 Heraklion-Crete, Greece.}
\affiliation{Department of Physics, University of Crete, 71003, Heraklion-Crete, Greece.}

\email{ptr@iesl.forth.gr}
\date{\today}

\begin{abstract}

We report the production of highly spin-polarized Deuterium atoms via photodissociation of deuterium iodide at 270 nm. The velocity distribution of both the deuterium and iodine photodissociation products is performed via velocity mapping slice-imaging. Additionally, the angular momentum polarization of the iodine products is studied using polarization-sensitive ionization schemes. The results are consistent with excitation of the $A^1\Pi_1$ state followed by adiabatic dissociation. The process produces $\sim$100\% electronically polarized deuterium atoms at the time of dissociation, which is then converted to $\sim 60\%$ nuclear D polarization after $\sim 1.6$ ns. These production times for hyperpolarized deuterium allow collision-limited densities of $\sim 10^{18}$ cm$^{-3}$, which is $\sim 10^6$ times higher than conventional (Stern-Gerlach separation) methods. We discuss how such high-density hyperpolarized deuterium atoms can be combined with laser fusion to measure polarized D-D fusion cross sections. 

\end{abstract}

\pacs{33.80.Gj, 32.10.Fn}
\maketitle

Controlling the nuclear spin polarization in fusion reactions offers important advantages, such as larger reaction cross sections, control over the emission direction of products, and in some cases eliminating hazardous neutron emission~\cite{Ciullo2016,Engels2014}. In the case of the five-nucleon reactions D + T $\rightarrow$ n + $^4$He and D + $^3$He $\rightarrow$ p + $^4$He, it is well known that the reaction cross section increases by $\sim$50\% when the fused nuclei have oriented nuclear spins~\cite{NucFusPolPart,Leeman71}. 

For the 4-nucleon D+D reaction, over the important energy range of 10-100 keV, the situation is unclear, since several predictions range from enhancement of the reaction, suppression of the reaction, or almost no effect at all~\cite{Fletcher94,Hale84,Hofman84,Lemaitre93,Zhang86,Zhang99}. The technical challenges of measuring fusion polarization dynamics limits the number of experiments which pursue this direction to two facility-scale experiments~\cite{Engels15}. These experiments achieve nuclear spin polarization by magnetically separating the nuclear spin states via the Stern-Gerlach effect in molecular beams, a process however which limits the achievable production rates to $\sim 10^{16}$ atoms s$^{-1}$~\cite{Engels2014}, and the projected D-D fusion reaction rate to as low as $\sim$0.01 neutrons s$^{-1}$~\cite{Schieck10}. Polarized solid deuterium targets can be produced where the densities can reach as high as $2.5\times 10^{19}$ spins/cm$^3$, albeit in much lower polarization close to 10\%~\cite{Iio04}. Another approach for the production of polarized sold fuel involves accumulating the hyperpolarized deuterium in specially coated storage cells~\cite{Engels15}. 

Laser photodissociation of hydrogen halides has been shown to produce highly polarized H atoms~\cite{Rakitzis03,RakitzisJCP04,SofikitisJCP08,SofikitisEPL08}. The photodissociation process initially polarizes the electron spin up to 100\% (for photofragment velocities parallel to the propagation direction of the circularly-polarized dissociation laser)~\cite{VanBrunt68}; due to the hyperfine interaction, the polarization oscillates back and forth between the electronic and the nuclear spin on the $\sim$1-ns timescale. By terminating this polarization exchange appropriately, for example by ionizing the atom, one can isolate the polarization on the electronic or nuclear spin. This ultrafast production mechanism allows the production of spin-polarized H atoms at densities many orders of magnitude higher than the traditional Stern-Gerlach spin-separation method.
This pulsed production of hyperpolarized H, if also extended to the production of hyperpolarized D atoms, can be used to study polarized D fusion reactions at high density, with much larger signals than conventional beam methods.  The D atoms can be used to produce pulsed ion beams, allowing the possibility of measuring polarized fusion reaction cross sections over a large energy range. The pulsed production can also be combined with Inertial Confinement Fusion (ICF), where fusion is induced by an intense short laser pulse. The photolysis and fusion pulses can be timed so that the D nuclei is maximally polarized when fusion occurs; such a combination can offer a straightforward method for measuring the effect of polarization in the D reactions, in densities close to $10^{18}$ cm$^{-3}$, at collision energies of $\sim 10 - 50$ keV. 

Theoretical calculations of the photofragment polarizations (over a range of UV photodissociation energies) have been performed for nearly all the hydrogen halides: HF/DF ~\cite{BalintKurti02}, HCl/DCl~\cite{BrownJPCA04}, HBr~\cite{Smolin06}, and HI/DI~\cite{BrownJCP05}.  At least two competing dissociation mechanisms occur: excitation to a dissociative state followed by adiabatic dissociation producing hydrogen atoms with spin-down electrons, excitation to another dissociative state followed by adiabatic dissociation producing hydrogen atoms with spin-up electrons (also nonadiabatic transitions between dissociative states can change the H-atom polarization). Therefore, depending on the contribution from each mechanism, the hydrogen electronic spin polarization $P_e$ can range from fully spin up ($P_e$=1) to fully spin-down ($P_e$=-1), and anything in between, including unpolarized ($P_e$=0). Calculations predict that $P_e$ can take values over this complete range, and depends strongly on the photolysis energy, on the halide cofragment (X=F,Cl,Br, or I), and the hydrogen isotope (H or D). For example, near maximal H polarizations ($\vert P_e \vert$ = 1) are calculated at some energies for photodissociation of the hydrogen halides HCl, HBr, and HI (of which HCl and HBr have been measured at 193 nm ~\cite{Rakitzis03,RakitzisJCP04}), and the deuterium halides DF and DI (which have not been measured previously). Here we measure the polarization of the D and I photofragments from the photodissociation of DI molecules at 270 nm, with the goal of demonstrating a source of highly polarized high-density D atoms.

The experiments are performed using a molecular beam set-up, to facilitate polarization sensitive detection. We measure the spin polarization of the iodine atoms via a Resonance-Enhanced Multiphoton Ionisation (REMPI) scheme and infer the deuterium polarization by angular momentum conservation; past experiments of direct measurements of hydrogen spin polarization~\cite{SofikitisJCP08,SofikitisEPL08} give excellent agreement with indirect detection. Measuring directly the spin polarization in hydrogen (or deuterium) atoms, usually requires the use of Vacuum-Ultra-Violet (VUV) sources~\cite{SofikitisJCP08,SofikitisEPL08,Broderick14}. Non VUV detection of spin polarized hydrogen has been reported, but using laser-induced fluorescence (LIF) detection and not REMPI~\cite{Bougas10}. 

The experimental set-up has been explained in detail elsewhere~\cite{Gebhardt01,Papadakis06}. Briefly, deuterium iodide is mixed with He in a 50\% ratio and expanded into the chamber via a pulsed nozzle, giving a molecule flux is $\sim 10^{17}$ molecules/s. The molecular beam is intersected at right angles roughly 50 cm downstream by the focused photolysis and probe laser beams. The resulting ion sphere is focused by a single ion lens on a gated position sensitive detector , thus performing velocity map imaging.~\cite{Eppink97}.  

\begin{figure}[t!]
	\centering
	\includegraphics[width=0.48\textwidth]{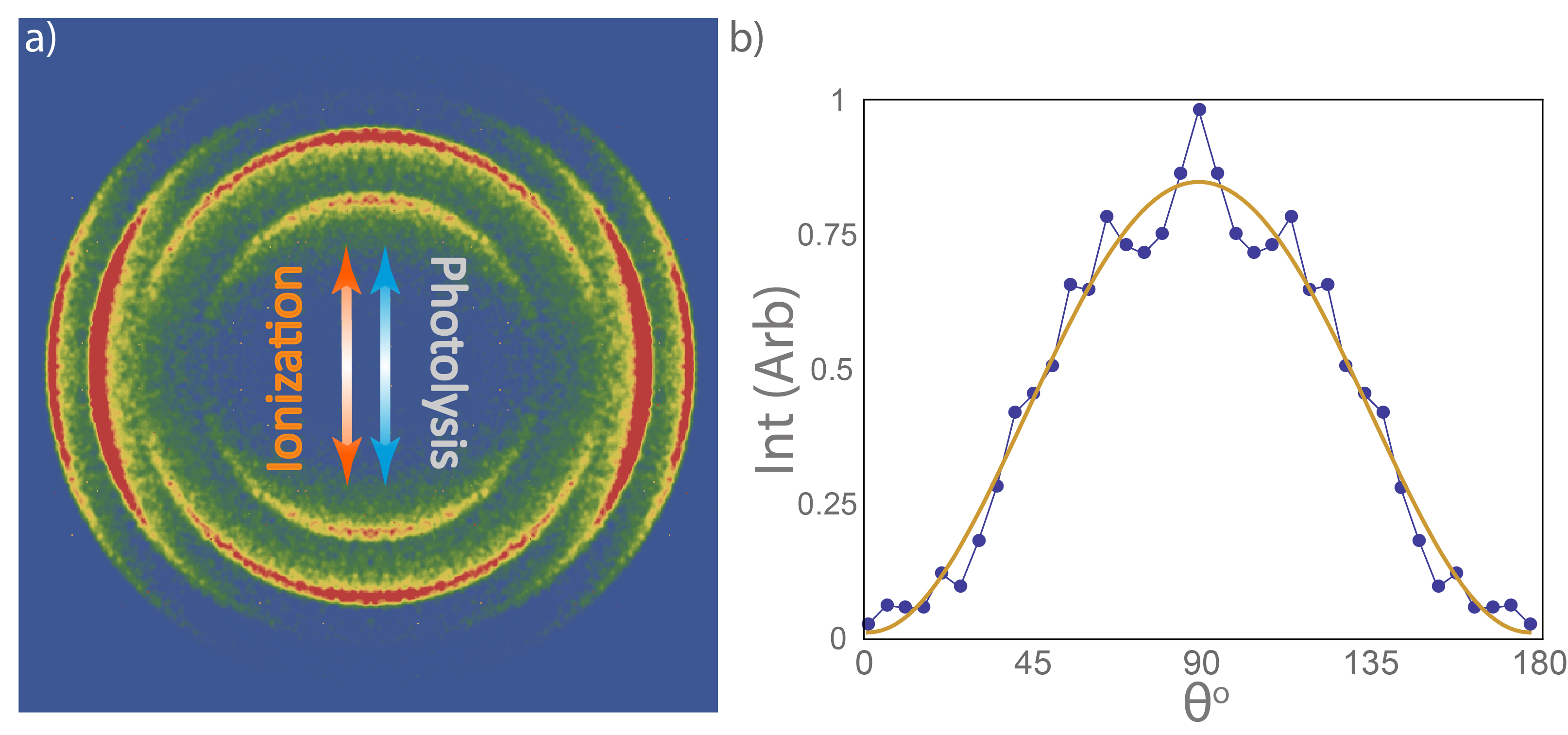}
	\caption
	{\footnotesize 
		a) Sliced ion image of the deuterium ions produced by the DI photodissociation. The direction of propagation of the photolysis laser is along the $90-270^o$ direction and opposite for the ionisation laser, while the direction of their linear polarization is marked by the arrows. b) Angular distribution of the ions corresponding to $DI+270 nm \rightarrow D+I(2P3/2)$ (points) and fit with Eq.~\ref{CircFitB} (orange line).}
	\label{D2Data}
\end{figure}

In Fig.~\ref{D2Data} we see a characteristic image of the D ions produced by dissociation of DI and subsequent ionisation of the D photofragments via 2+1 REMPI at 243.16 nm. From inner to outer, the rings are assigned as:  (1) DI + 270 nm $\rightarrow$ D + I*($^2P_{1/2}$) ($v_D$ = 7242 m/s), (2) DI + 243 nm $\rightarrow$ D + I*($^2P_{1/2}$) ($v_D$ = 10184 m/s), (3) DI+270 nm $\rightarrow$ D+I($^2P_{3/2}$)  ($v_D$ = 12032 m/s), and (4) DI + 243 nm $\rightarrow$ D+I($^2P_{3/2}$) ($v_D$ = 13880 m/s), where $v_D$ is the measured velocity of the D atoms. Notice that the deuterium atoms correlated to the production of I* atoms (two inner rings) have completely different angular distributions to those correlated to the production of I atoms (two outer rings), which in combination with the high velocity of the photoproducts leads to their spatial separation in a few ns assuming a laser focus of several tens of $\mu$m. Note that the ions produced by 243 nm dissociation are not produced when laser detection is not used.

The angular distribution of the polarized photofragments can be expressed as an even expansion of Legendre polynomials~\cite{Choi86,ZareIEEE63,Siebbeles94} up to the fourth order:
\begin{equation}\label{beta2}
I(\theta)/I_0 = 1+\beta_2 P_2(cos\theta) + \beta_4 P_4(cos\theta)
\end{equation}
where $P_n(x)$ is the nth-order Legendre polynomial. In Fig.\ref{D2Data}.b we see the angular distribution of the deuterium ions corresponding to the production of $I(^2P_{3/2})$ atoms at 270 nm, as well as a fit using Eq.\ref{beta2}. In this case, as there is no detection sensitivity to the D-atom spin, $\beta_2=\beta$ and $\beta_4=0$. The value of the spatial-anisotropy $\beta$ parameter extracted by this fit is $\beta = -0.98 \pm 0.03$, corresponding to an almost purely perpendicular transition, in agreement with theoretical predictions~\cite{BrownJCP05}. 

The total intensity $I_0$ of the iodine photofragments, and the $\beta_2$ coefficient, describing the angular distribution, are expressed in terms of the spatial anisotropy $\beta$ and a linear combination of the $a_q^{(1)}(p)$ parameters~\cite{Zhang97}:
\begin{equation}\label{{CircFitI}}
I_0 = 1 + \frac{s_1}{3} \Big[ \Big(1 - \frac{\beta}{2}\Big) a_0^{(1)}(\perp) + \sqrt{2} Re[a_1^{(1)}]) \Big]
\end{equation}
\begin{equation}\label{CircFitB}
\beta_2 = \frac{1}{I_0} \Big\{ \frac{2 s_1}{3} \Big[ \Big(1- \frac{\beta}{2}\Big)a_0^{(1)} (\perp) - \frac{\sqrt{2}}{2} Re[a_1^{(1)}])  \Big] -\frac{\beta}{2}\Big\}
\end{equation}
where $s_1$ is the sensitivity factor of the REMPI transition to the $a_q^{(1)}(p)$ parameters, given by $s_1=-\sqrt{3}\langle G^{(1)}\rangle$ for RCP and $s_1=+\sqrt{3}\langle G^{(1)}\rangle$ for LCP probe light, and $\langle G^{(1)}\rangle \approx 0.400$ is the time-averaged k=1 hyperfine depolarization coefficient due to the iodine nuclear spin (I = 5/2). Thus, the only way that the intensity $I_0$ and the $\beta_2$ can differ upon reversal of the probe (or photodissociation) laser polarization helicity, is for the iodine atoms to be polarized (i.e. the $a_q^{(1)}(p)$ parameters are not zero). The $a_0^{(1)}(\perp)$ parameter is proportional to the j projection $\langle m_j(I) \rangle	 = \sqrt{j(j+1)}a_0^{(1)}(\perp)$, with j=3/2, while the $Re[a_1^{(1)}]$ parameter is related to interference from multiple dissociating states accessed by parallel and perpendicular transitions.  

\begin{figure}[t!]
	\centering
	\includegraphics[width=0.5\textwidth]{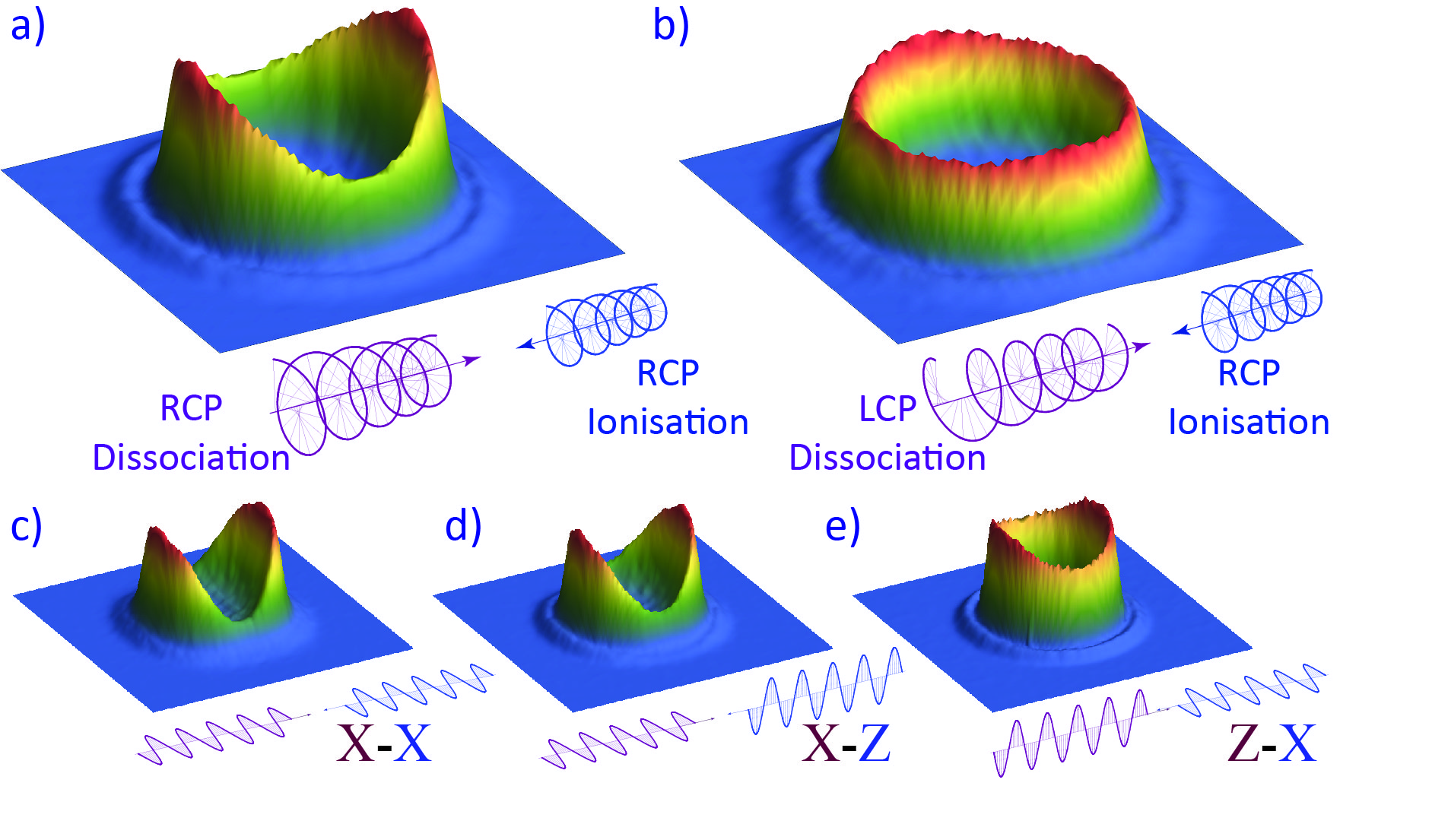}
	\caption
	{\footnotesize 
		a) Sliced ion image of I ions produced with both the photolysis and ionisation lasers right circularly polarized (RCP) b) Sliced ion image of I ions produced the ionisation laser RCP and the photolysis laser left circularly polarized (LCP) c) Sliced ion image with the photolysis and the ionisation lasers linearly polarized parallel to the image plane (X) d) Sliced ion image with the photolysis laser linearly polarized along X, and the ionisation laser linearly polarized perpendicular to the image plane (Z) e) Sliced ion image with the photolysis laser linearly polarized along Z, and the ionisation laser linearly polarized along X.}
	\label{IodineData}
\end{figure}

In Figs \ref{IodineData}a and \ref{IodineData}b we see two sliced ion images of iodine  atoms produced by the DI+270 nm  $\rightarrow D+I(^2P_{3/2})$ and detected via the $^2P_{3/2} \rightarrow \rightarrow ^4P_{1/2}$ REMPI transition at 303.6 nm. In Fig.\ref{IodineData}a, both the photolysis and the ionisation lasers are right-circularly polarized (RCP); the angular distribution is strongly anisotropic, with most of the photofragments recoiling preferentially along the propagation direction of the laser beams. For Fig.\ref{IodineData}b, the photolysis laser has opposite helicity: left-circularly polarized (LCP), and we see a very big change in the angular distribution, as it is nearly isotropic. The ratio of the total intensities of the images in Figs.3a and 3b is $I_0(RR)/I_0(LR) = 1.8\pm0.2$. 

Figures~\ref{IodineData}c-e show images using linearly polarized photolysis and ionization laser light, aligned in the geometries (XX), (XZ), and (ZX), respectively (see Fig.\ref{IodineData} for details). The $\beta_2$ anisotropy parameter of the corresponding geometries (determined from fitting the angular distributions using Eq.~\ref{beta2}) can be related to the $\beta$, $a_0^{(2)}$ and $a_2^{(2)}$ parameters as:
\begin{equation}\label{XX} 
\beta_2^{XX}=\frac{\beta+\frac{s_2}{7} \Big(5 a_0^{(2)}(\perp)-4\sqrt{6}a_2^{(2)}(\perp) \Big)}{1-\frac{s_2}{5}\Big(a_0^{(2)}(\perp)+2\sqrt{6}a_2^{(2)}(\perp) \Big)}
\end{equation}
\begin{equation}\label{XZ} 
\beta_2^{XZ}=\frac{\beta+\frac{s_2}{2} \Big( a_0^{(2)}(\perp)+\sqrt{6}a_2^{(2)}(\perp) \Big)}{1-\frac{s_2}{2}\Big(a_0^{(2)}(\perp)+\sqrt{6}a_2^{(2)}(\perp) \Big)}
\end{equation}
\begin{equation}\label{ZX}
\beta_2^{ZX}=\frac{\frac{s_2}{3} \Big(3 a_0^{(2)}(\perp)+\sqrt{6}a_2^{(2)}(\perp) \Big)}{1-\frac{s_2}{3}\sqrt{6}a_2^{(2)}(\perp)}
\end{equation}
where $s_2=(-5/4)\langle G^{(2)}\rangle$, and $\langle G^{(2)} \rangle \approx 0.233$ is the average k=2 hyperfine depolarization coefficient. Fitting images in the geometries shown in Fig.\ref{IodineData}c-e with equations \ref{XX}-\ref{ZX} yields the $a_0^{(2)}(\perp)$  and $a_2^{(2)}(\perp)$ polarization parameters; the $a_0^{(2)}(\perp)$  gives an independent measure of $\langle m_j(I) \rangle$, as $a_0^{(2)}(\perp)=\langle \frac{3 m_j(I)^2}{j(j+1)}-1\rangle$, and $a_2^{(2)}(\perp)$ gives the interference between perpendicular transitions excited with opposite helicity.

\begin{figure}[t!]
	\centering
	\includegraphics[width=0.45\textwidth]{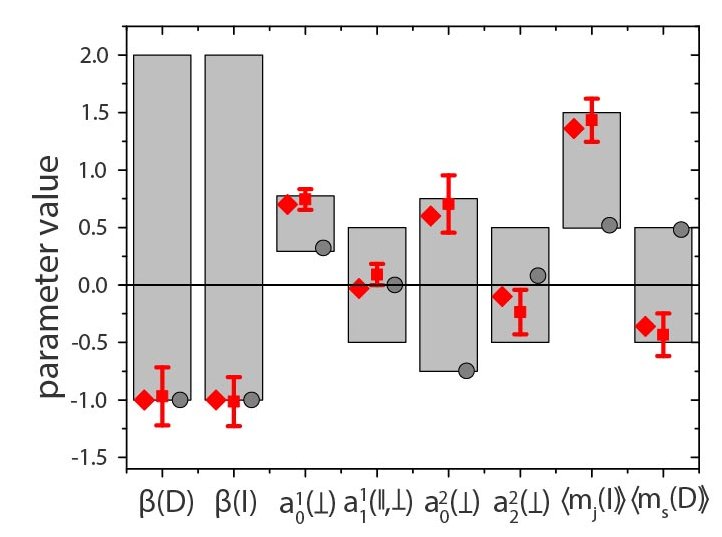}
	\caption
	{\footnotesize 
		Results for the β parameter for the D and I($^2P_{3/2}$) atoms from the photodissociation of DI at 270 nm, and the I($^2P_{3/2}$) polarization parameters $a_q^{(k)}(p)$, which give $\langle m_j(I) \rangle \approx 1.5$ and $\langle m_s(D) \rangle \approx -0.5$. Error bars are $2\sigma$ confidence intervals, and the grey bars give the allowed range of the parameters. Theoretical predictions [23] are given for energies of 47,000 $cm^{-1}$ (red diamonds) and 37,000 $cm^{-1}$ (grey circles).}
	\label{Results}
\end{figure}

\begin{figure}[t!]
	\centering
	\includegraphics[width=0.45\textwidth]{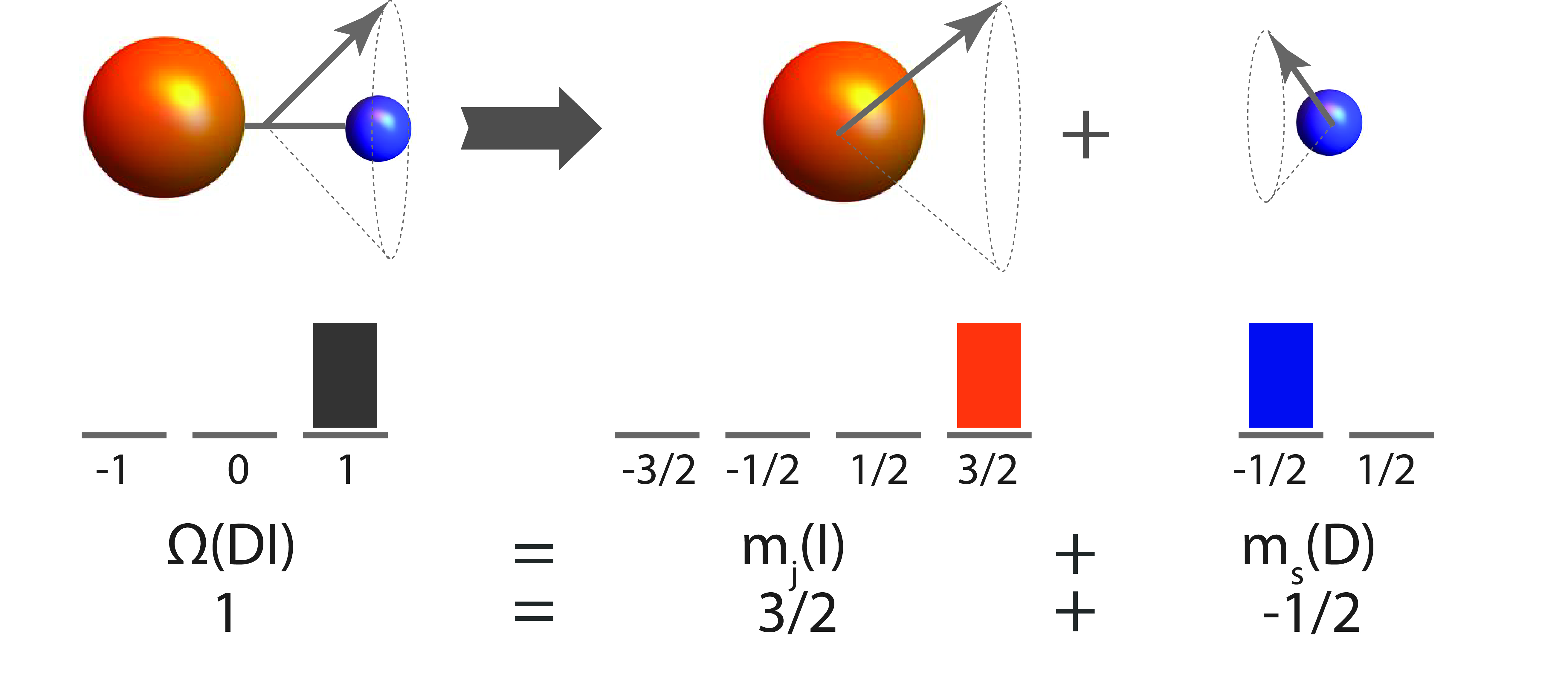}
	\caption
	{\footnotesize 
		Angular momentum projection of $\Omega$=+1 is prepared for the DI molecules from the circularly polarized photolysis beam, and is distributed to the angular momentum projections of the photofragments after }
	\label{Momenta}
\end{figure}

In Fig.\ref{Results} we present graphically the fitted values of the $\beta$ parameter, for the D and I($^2P_{3/2}$) photofragments from the photodissociation of DI at 270 nm, as well as the $a_q^{(k)}(p)$ parameters for $I(^2P_{3/2})$, along with the corresponding values of $\langle m_j(I) \rangle$ and the inferred D-atom electron spin-projection $\langle m_s(D) \rangle = 1-\langle m_j(I) \rangle$. The gray histograms represent the physical ranges of these parameters. The value of each of the $a_0^{(1)}(\perp)$ and $a_0^{(2)}(\perp)$  parameters is maximal, meaning that $\langle m_j(I) \rangle = 1.5$ is also maximal, constraining the D photofragment electron spin polarization to be maximally polarized in the opposite direction, $\langle m_s(D) \rangle = -0.5$~\cite{BrownJCP05}, as shown in Fig.~\ref{Momenta}. These m-state values are explained by excitation of the $A^1\Pi_1$ state at 270 nm, followed by adiabatic dissociation:
\begin{equation}\label{ReactionEq}
DI(\Omega_A=\pm 1)\rightarrow D(m_s = \mp 1/2) + I(m_j = \pm 3/2)
\end{equation}

The measured $\beta$ and polarization parameters are consistent with those of the dissociation mechanism predicted at a significantly higher energy, $\sim47,000\, cm^{-1}$~\cite{BrownJCP05}. However, at lower energies, the dissociation is predicted to switch to excitation and adiabatic dissociation via the $a^3\Pi_1$ state, producing $m_j(I)=+1/2$ and $m_s(D)=+1/2$~\cite{BrownJCP05}; the energy of this mechanism switch is strongly model dependent. Our measurements here at $\sim 37,000\, cm^{-1}$ do not observe this mechanism switch, indicating that the dissociation models may need to be modified.

We note that Eq.~\ref{ReactionEq} applies exactly only for photofragment recoil direction \textit{\^{v}} parallel to the photolysis polarization direction \textit{\^{k}}; the photofragment polarization falls as \textit{\^{v}} $\cdot$ \textit{\^{k}} = cos$\theta_{vk}$. This loss in polarization can be minimized by preferentially aligning the HX bonds along \textit{\^{k}} just before photodissociation, using nonresonant molecular alignment with a strong laser pulse~\cite{Friedrich95,Stapelfeldt03,Vrakking97}. 

\begin{figure}[h!]
	\centering
	\includegraphics[width=0.45\textwidth]{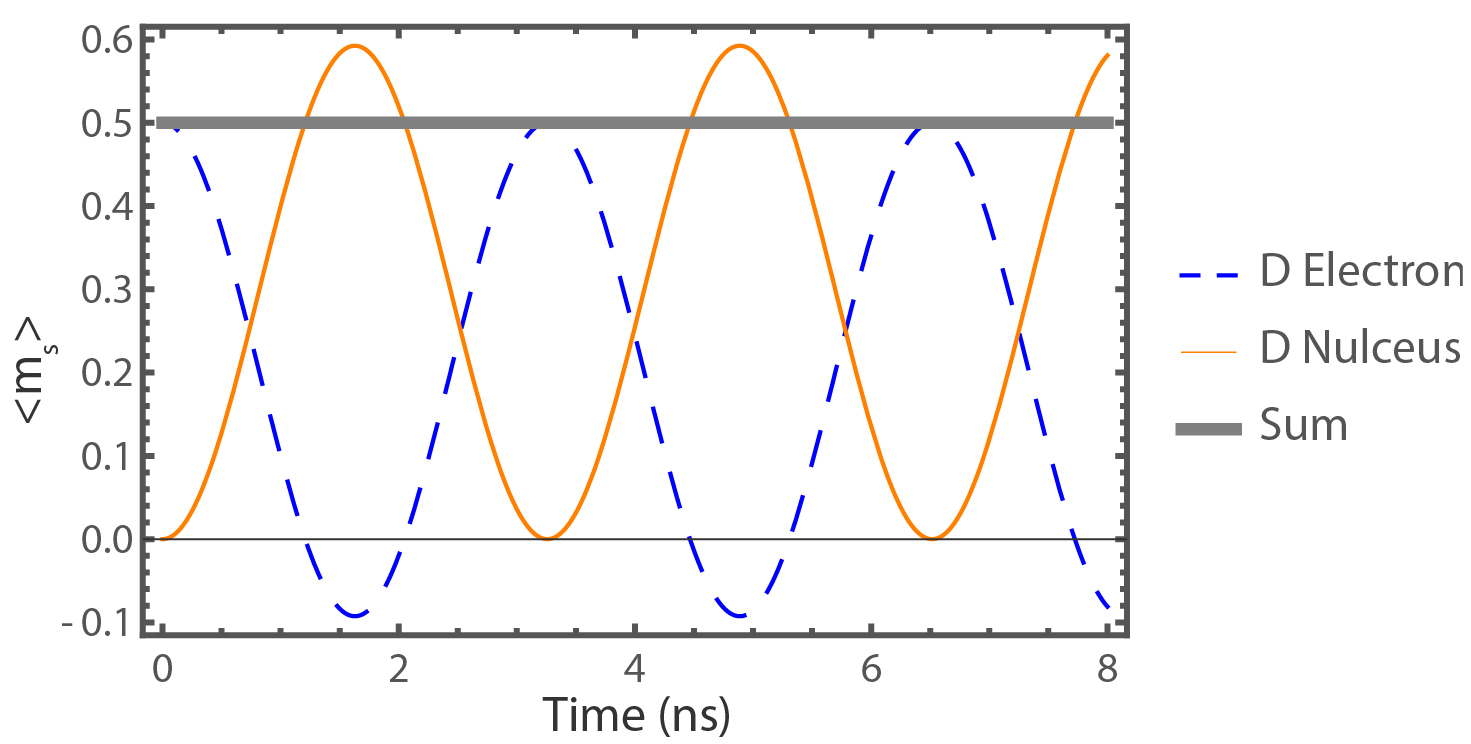}
	\caption
	{\footnotesize 
		Evolution of the spin projection expectation values of the electron (dashed line), the nucleus (thin solid line) and sum (thick solid line) due to the hyperfine interaction. }
	\label{DepolBeat}
\end{figure}

The photodissociation process initially polarizes the electron spin S leaving the nuclear spin I initially unpolarized. The D atoms are found in a coherent superposition of the total angular momentum states $\vert F, M_F \rangle$, defined by the coupling F= S + I, which are the eigenstates of the system. Therefore, the system evolves in time, transferring the polarization of the electron spin to the nuclear spin, and back~\cite{Altkorn85}. Such a polarization transfer has been experimentally demonstrated in several cases~\cite{SofikitisHFPexp07,Bartlett07}. Here, the polarization transfer can be quantified~\cite{Rubio-Lago06} using Eqs. 3-6 in Ref.~\cite{Rakitzis05}, giving:
\begin{subequations}
\begin{eqnarray}\label{beatingEq1}
m_s(D)=\frac{16}{27} sin^2(\frac{\Delta E}{2\hbar} t)
\\\label{beatingEq2}
m_j(I)=\frac{1}{2}-\frac{16}{27} sin^2(\frac{\Delta E}{2\hbar} t)
\end{eqnarray}
\end{subequations}
where $\Delta E=327.37 MHz$ is the hyperfine splitting in the deuterium atom~\cite{Nafe47}. We show the polarization dependence of the D electron and nuclear spin, by plotting Eqs.~\ref{beatingEq1} and ~\ref{beatingEq2} in Fig.\ref{DepolBeat}. At t=0, the electron spin is fully polarized, and the nuclear spin is unpolarized. The electron spin polarization is transferred to the nuclear spin and back, with a period of $\sim$3.2 ns. An interesting aspect of these dynamics is that there exist times (for example around 1.6 and 4.8 ns) where the electron spin polarization is opposite to the initial one, albeit with a much smaller value. At the same times, the spin projection of the D nucleus reaches a value of ~60\%.

The main depolarizer of the D atoms will be the halogen-atom cofragment, here $I(^2P_{3/2})$. Thermal spin-exchange rates of H/D with halogen atoms are not available, but are of order $10^{-9} cm^3s^{-1}$ for collisions with alkali atoms~\cite{Anderson95}. Using this rate, maximal densities of $\sim 10^{18} cm^{-3}$ of spin-polarized D atoms can be produced over the needed time of ~1 ns. However, for the photodissociation of DF near 193 nm, it is predicted that both the D and the F$(^2P_{3/2})$ cofragment will have the same polarization: $\langle m_s(D)\rangle = \langle m_j(F)\rangle =0.5$; there is even the possibility of polarizing the F($^2P_{3/2}$) further via infrared excitation of the DF before photodissociation~\cite{Altkorn85,SofikitisHFPexp07,Bartlett07,Rubio-Lago06,Rakitzis05}. As all colliding species will have similar polarizations, the depolarization rate of D atoms should be significantly reduced, and perhaps densities of greater than $\sim 10^{18}\,cm^{-3}$ spin-polarized D atoms can be produced.

Calculations using modified MEDUSA code~\cite{Christiansen74} have shown that the irradiation of $10^{18}\,cm^{-3}$ spin-polarized D atoms with a 6 kJ (with $\lambda \sim 1\,\mu m$), 1 ps laser pulse~\cite{Ayers12,Haefner11}, focused to $10\,\mu m,$ will heat the resulting ions to average collision (thermal) energies of $\sim$10 keV, producing $\sim$100 neutrons/pulse from D-D fusion reactions. Past experiments have demonstrated a neutron production in excess of $10^6$ per pulse using a 62 J, 1 ps laser pulse, focused to $20\,\mu m$, and D$_2$ gas densities of up to $\sim 10^{20}\, cm^{-3}$~\cite{Fritzler02}. These results show that the study of D-D fusion with high signals is possible, using high-density polarized D atoms and laser-initiated fusion.

The experimental work was conducted at the Ultraviolet Laser Facility at FORTH-IESL, supported in part by the EC H2020 project LASERLAB-EUROPE (Grant Agreement No. 654148), and the European Research Council (ERC) Grant TRICEPS (GA No. 207542).

\bibliographystyle{apsrev}

\bibliography{PolarizedFusionRefs} 

\end{document}